\documentclass[twocolumn]{jpsj2} %% two-column layout
\usepackage{bm}

\title{Metal-Insulator Transition and Magnetic Order in the Pyrochlore Oxide Hg$_2$Ru$_2$O$_7$}

\author{Makoto \textsc{Yoshida}\thanks{E-mail address: yopida@issp.u-tokyo.ac.jp}, 
Masashi \textsc{Takigawa}\thanks{E-mail address: masashi@issp.u-tokyo.ac.jp}, 
Ayako \textsc{Yamamoto}$^{1}$, 
and Hidenori \textsc{Takagi}$^{1,2}$}

\inst{Institute for Solid State Physics, University of Tokyo, Kashiwa, Chiba 277-8581 \\ 
$^{1}$RIKEN Advanced Science Institute, Wako, Saitama 351-0198\\
$^{2}$Department of Advanced Materials, University of Tokyo, Kashiwa, Chiba 277-8561}

\abst{We report results of NMR experiments on the ruthenium oxide 
Hg$_2$Ru$_2$O$_7$ with the pyrochlore structure, 
which exhibits a metal-insulator transition at 
$T_{\rm MI}$ = 107 K. In the metallic phase above $T_{\rm MI}$, 
the nuclear spin-lattice relaxation rate 1/$T_1$ and the Knight shift at 
the Hg sites follow the Korringa relation, indicating the absence 
of substantial spatial spin correlation. At low temperatures in the 
insulating phase, $^{99,101}$Ru-NMR signals are observed at zero 
magnetic field, providing evidence for a commensurate antiferromagnetic 
order. The estimated ordered moment is about 1~$\mu _B$ per Ru, much 
smaller than 3 $\mu _B$ expected for the ionic (4$d$)$^{3}$ 
configuration of Ru$^{5+}$. Thus, the localized spin models are 
not appropriate for the insulating phase of Hg$_2$Ru$_2$O$_7$. 
We also discuss possible antiferromagnetic spin structures.}

\kword{pyrochlore structure, ruthenium oxide, Hg$_2$Ru$_2$O$_7$, NMR, metal-insulator transition, magnetic order}

\begin{document}
\maketitle

\section{Introduction} 
\label{introduction}

The pyrochlore lattice, a network of corner-sharing tetrahedra, 
is known for strong geometrical frustration\cite{Gardner}. 
Antiferromagnetically coupled localized spins on this lattice exhibit 
a massive degeneracy of low-energy states.  Both 
classical\cite{Reimers921,Moessner981,Moessner982} 
and quantum\cite{Canals981,Canals001,Tsunetsugu} spin systems 
on the pyrochlore lattice with the nearest neighbor Heisenberg interaction 
are believed to remain disordered down to zero temperature 
with spin liquid ground states. Real materials, however, have secondary 
interactions, which may stabilize a magnetic order. 
The effects of secondary interactions such as anisotropy\cite{Bramwell}, 
longer range Heisenberg\cite{Reimers921} or dipolar\cite{Palmer} interactions, 
Dzyaloshinsky-Moriya (DM) interaction\cite{Elhajal}, and spin-lattice 
coupling\cite{Yamashita,Tchernyshyov}have been studied theoretically. 

Although the concept of frustration is well established for localized 
spins, it is not yet clear for itinerant electron systems. Nevertheless, 
some anomalous properties of itinerant electrons on a pyrochlore 
lattice are considered to be related to geometrical frustration. A 
well-known example is the spinel oxide LiV$_2$O$_4$, where 
mixed valent V atoms form a pyrochlore lattice. This compound 
shows heavy-electron behavior such as extremely enhanced specific 
heat and magnetic susceptibility at low temperatures\cite{Kondo}. 
The origin could be the large degeneracy of low-lying states 
inherent to frustrated systems, and the mechanism of the heavy-electron 
behavior is probably different  from the Kondo resonance in 
$f$-electron systems\cite{Urano}. 

Some pyrochlore oxides show metal-insulator (MI) transitions by 
changing temperature or pressure.  Such materials may help us fill 
the gap of our understanding of the frustration effects of localized spins and 
itinerant electrons.  The MI transition in Cd$_2$Os$_2$O$_7$ 
at $T_{\rm MI}$ = 226~K is continuous with no change 
in structural symmetry\cite{Sleight, Mundrus}. 
The Slater mechanism was proposed to explain the transition, i.e., unit 
cell doubling due to an antiferromagnetic order below $T_{\rm MI}$ 
opens a gap at the Fermi level.  In another example, 
Tl$_2$Ru$_2$O$_7$, the discontinuous MI transition is accompanied 
by a pronounced structural change at 120 K \cite{Takeda,Sakai,Lee}. 
The low-temperature insulating phase has a spin-singlet ground state 
with a gap in magnetic excitations\cite{Sakai}. Lee \textit{et al.} 
proposed that orbital ordering among 4$d$ electrons of Ru$^{4+}$ ($4d^4$) 
ions stabilized by structural distortion results in the formation of effectively 
one-dimensional spin chains and interpreted the excitation gap as the 
Haldane gap of spin 1 Heisenberg chains\cite{Lee}. 

Recently, a MI transition has been reported in another Ru-pyrochlore 
oxide, Hg$_2$Ru$_2$O$_7$\cite{Yamamoto,Klein,Takeshita}. This 
compound shows a first-order MI transition with a hysteresis at 
$T_{\rm MI}$ = 107 K. Upon cooling through $T_{\rm MI}$, 
the resistivity increases abruptly and the crystal structure changes 
from cubic to one of lower symmetry, although the structural refinement has not been 
performed yet in the low-temperature phase\cite{Yamamoto}. In the metallic phase 
above $T_{\rm MI}$, the magnetic susceptibility $\chi $ increases 
slightly with decreasing temperature but decreases discontinuously  
below $T_{\rm MI}$, suggesting either an antiferromagnetic order\cite{Klein} 
or a spin-singlet ground state\cite{Yamamoto}. Although such behavior 
is similar to Tl$_2$Ru$_2$O$_7$, the valence of Ru in 
Hg$_2$Ru$_2$O$_7$ is 5+ with the $4d^3$ configuration and, 
in the localized picture, we expect  a half filled $t_{2g}$ manifold with 
spin 3/2 and no orbital degree of freedom.  Therefore, the magnetic ground 
state should be quite different from Tl$_2$Ru$_2$O$_7$. 
   
In this paper, we report results of NMR measurements in 
Hg$_2$Ru$_2$O$_7$ and discuss the microscopic magnetic properties. 
We found from $^{199}$Hg-NMR that the metallic phase above 
$T_{\rm MI}$ does not exhibit substantial spatial magnetic correlation. 
$^{99,101}$Ru-NMR lines are observed at zero field at low 
temperatures, providing evidence for an antiferromagnetic 
order in the insulating phase. The ordered moments are 
estimated to be about 1 $\mu _B$ per Ru. The small moment compared 
with the expected value (3 $\mu_B$) from Ru$^{5+}$ ($S$ = 3/2) 
implies that the localized spin model with strong Hund's coupling is 
inadequate to explain the magnetic properties of Hg$_2$Ru$_2$O$_7$. 
   
\section{Experiment} 

Powder samples of Hg$_2$Ru$_2$O$_7$ were prepared as 
described in ref. \citen{Yamamoto}. Most of the NMR measurements 
were performed on sample $A$ with natural 
isotopic abundance. We also prepared isotopically enriched sample $B$ 
containing 97.7\% $^{99}$Ru for the unambiguous site assignment of NMR lines 
at zero magnetic field. For sample $B$, we first prepared RuO$_2$ by heating 
isotopically enriched Ru metal at 900 $^{\circ }$C in air, followed by the same 
procedure used for sample $A$. 
Most of the NMR spectra of $^{199}$Hg nuclei were 
obtained by summing up the Fourier transform of the spin-echo signal at 
equally spaced frequencies with a fixed magnetic field~\cite{Clarke951}. 
The NMR spectra of $^{99, 101}$Ru nuclei at zero magnetic field were obtained 
by recording the integrated intensity of the spin-echo signal at discrete 
frequencies.  The nuclear spin-lattice relaxation rate 1/$T_1$ was measured 
by the inversion recovery method. Various properties of nuclei relevant to 
the present study are summarized in Table I. 
\begin{table}[b]
\caption{Nuclear spin $I$, gyromagnetic ratio $\gamma _n$, 
quadrupole moment $Q$, and natural abundance for $^{199}$Hg and $^{99,101}$Ru.}
\label{nuclei}
\begin{tabular}{ccccc}
\hline
  & $I$ (MHz) & $\gamma _n$ (MHz/T) & $Q$ (barn) & abundance (\%) \\
\cline{2-5}
$^{199}$Hg & 1/2 & 7.6258 & -- & 16.9\\
$^{99}$Ru & 5/2 & 1.9607 & 0.079 & 12.8\\
$^{101}$Ru & 5/2 & 2.1975 & 0.457 & 17.1\\
\hline
\end{tabular}
\end{table} 

\section{Results and Discussion} 

In $\S$\ref{HgNMR}, we present the results of $^{199}$Hg-NMR, which 
indicate that Hg$_2$Ru$_2$O$_7$ above $T_{\rm MI}$ is a correlated metal 
with negligible spatial spin correlation.  We also discuss the change in spectral 
shape and the loss of signal intensity across $T_{\rm MI}$, pointing to a magnetic order 
in the insulating phase.  In $\S$\ref{RuNMR}, we show the results of zero-field 
$^{99,101}$Ru-NMR at low temperatures, providing direct evidence for an 
antiferromagnetic order below $T_{\rm MI}$.  Possible magnetic structures 
are discussed in $\S$\ref{structure}.

\subsection{$^{199}$Hg-NMR} 
\label{HgNMR}

\begin{figure}[tb]
\begin{center}
\includegraphics[width=0.9\linewidth]{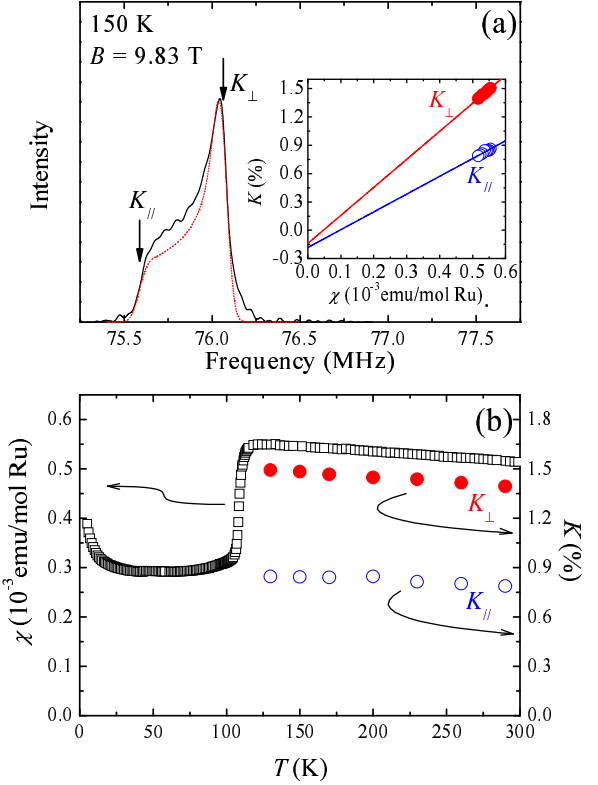}
\end{center}
\caption{(Color online) (a) $^{199}$Hg-NMR spectrum in the 
metallic phase ($T = 150$~K) obtained for sample $A$ in 
the magnetic field of 9.83~T ($^{199}\gamma B$ = 
74.953~MHz). The red dotted 
line shows the fit to the powder pattern for an axially symmetric 
Knight shift. The edge and peak of the spectrum marked by 
$K_{\parallel }$ and $K_{\perp}$ respectively correspond to the resonance 
frequencies for the field parallel and perpendicular to the trigonal 
$\langle 111\rangle$ axis. In the inset, $K_{\parallel}$ 
and $K_{\perp}$ are plotted against the magnetic susceptibility
$\chi$ (the $K-\chi$ plot) and fit to straight lines. (b) Temperature 
dependences of $\chi$ (open squares), $K_{\parallel }$ (open 
circles), and $K_{\perp}$ (solid circles).}
\label{fig1}
\end{figure}

The NMR frequency $\nu _r$ of $^{199}$Hg nuclei, which have spin 1/2 
and no quadrupole moment, is expressed in the paramagnetic state as, 
\begin{equation}
\nu _r =~^{199}\gamma B (1+K),
\label{Knightshift}
\end{equation} 
where $^{199}\gamma$ is the gyromagnetic ratio of $^{199}$Hg nuclei and 
$B$ is the external magnetic field. The Knight shift $K$ represents the shift in 
resonance frequency due to a magnetic hyperfine field induced by 
the external field. Since the Hg sites in the pyrochlore structure 
have trigonal symmetry $\bar 3m$, the Knight shift should follow a uniaxial 
angular dependence, 
\begin{equation}
K(\theta) = K_{\rm iso} + K_{\rm ax} \left(3 \cos^2\theta-1 \right), 
\label{KisoKax}
\end{equation} 
where $\theta$ is the angle between the local symmetry axis along the $\langle 111\rangle$ 
direction and the external field. For a powder sample, the NMR line shape represents 
the histogram $P(K)$ of the angular distribution of $K(\theta)$, 
\begin{equation}
P(K) \propto \left| \frac{d K}{d \cos\theta} \right| ^{-1}\propto |K-K_{\perp }|^{-1/2},
\label{PowderPattern}
\end{equation} 
for $K_{\parallel } \le K \le K_{\perp }$ or $K_{\perp} \le K \le K_{\parallel}$ 
and $P(K) = 0$ otherwise. Here, $K_{\parallel}= K_{\rm iso}+ 2K_{\rm ax}$ 
and $K_{\perp}= K_{\rm iso}- K_{\rm ax}$ are the Knight shifts for a field 
parallel and perpendicular to the symmetry axis, respectively. 

Figure 1(a) shows a typical $^{199}$Hg-NMR spectrum in the 
metallic phase above $T_{\rm MI}$.  The spectra can be fit 
reasonably well to eq.~(\ref{PowderPattern}) convoluted with 
a gaussian broadening function as shown by the red dotted line. Figure 1(b) 
shows the temperature dependences of $K_{\parallel}$ and $K_{\perp }$,  
corresponding to the edge and peak positions of the spectrum, respectively, 
and the magnetic susceptibility $\chi$ of the same sample. 
Both the Knight shifts and susceptibility increase slightly with decreasing 
temperature. 

The magnetic susceptibility $\chi$ is generally expressed as 
\begin{equation}
\chi = \chi_0 + \chi_{\rm spin}(T), 
\label{chi}
\end{equation} 
where the first term is the sum of the diamagnetic and orbital (van-Vleck) contributions, 
which are independent of temperature, and the second term is the spin contribution, 
which may depend on temperature. Similarly,
\begin{equation}
K_{\epsilon} = K_{\epsilon, 0} + K_{\epsilon, \rm spin}(T) \  (\epsilon  =~\parallel  {\rm or}  \perp),
\label{shift}
\end{equation} 
where the first (second) term is the chemical (spin) shift. 
The spin part of the susceptibility 
(per mole of Ru) and the spin Knight shift are linearly related as 
\begin{equation}
K_{\epsilon, \rm spin} = A_{\epsilon}^{\rm Hg}\frac{\chi_{\rm spin}}{N_A \mu_{B}}, 
\label{coupling}
\end{equation} 
where $A_{\epsilon}^{\rm Hg}$ is the hyperfine coupling between the 
conduction electron spin and the $^{199}$Hg nuclei representing the 
magnetic hyperfine field produced by a uniform magnetization of 1~$\mu_{B}$ 
per Ru atom. 

In the inset of Fig.~1(a), $K_{\parallel}$ and $K_{\perp}$ are plotted 
against $\chi$ (the $K$-$\chi$ plot) and fit to straight lines.  
The slopes of the lines give the hyperfine coupling constants 
$A_{\parallel}^{\rm Hg} = 10.5$, $A_{\perp}^{\rm Hg} = 16.6$ (T/$\mu _B$) 
or equivalently $A_{\rm iso}^{\rm Hg} = 14.6$,  $A_{\rm ax}^{\rm Hg} = -2.03$ (T/$\mu _B$). 
Thus, the hyperfine interaction is mainly isotropic, indicating that the major source of 
the hyperfine field at the $^{199}$Hg nuclei is the spin polarization on the Hg-$s$ orbital 
hybridized with the conduction band. 
The chemical shift of $^{199}$Hg nuclei is negative for most 
diamagnetic compounds and its absolute value is less than 
0.3\%\cite{Harris}, when dimethil-mercury (HgMe$_2$) is used as the 
reference material.  The value of $^{199}\gamma$ in Table I is also determined 
against HgMe$_2$ as the reference.  Since this value is much smaller than the
 observed values of $K_{\rm iso}$, we set $K_{\epsilon, 0}=0$ in eq.~(\ref{shift}).  
The $K-\chi$ plot in the inset of Fig.~1(a) then indicates that $\chi \sim \chi_{\rm spin}$. 

\begin{figure}[tb]
\begin{center}
\includegraphics[width=0.9\linewidth]{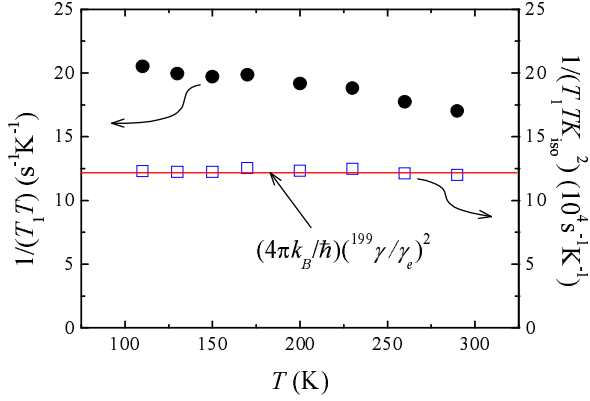}
\end{center}
\caption{(Color online) Temperature dependences of $1/(T_1T)$ (solid circles) 
and $1/(T_1TK_i^2)$ (open squares) at the Hg sites. The relaxation rate $1/T_1$ was 
measured at the peak frequency of the spectra in the field of 9.83 T. 
The solid line represents $S \equiv (4\pi k_B/\hbar )(^{199}\gamma /\gamma _e)^2$.}
\label{fig2}
\end{figure} 
We next discuss the nuclear relaxation rate.  For non-interacting electrons, 
$K_{\rm spin}$ and $1/(T_1T)$ (the nuclear spin-lattice relaxation rate 
$1/T_1$ divided by $T$) are basically temperature-independent 
and satisfy the following Korringa relation when the hyperfine field 
is due to $s$-electrons: 
\begin{equation}
\frac{1}{T_{1}TK_{\rm spin}^{2}} = \frac{4\pi k_{B}}{\hbar}\left(\frac{\gamma_{n}}{\gamma_{e}}\right)^{2}\equiv S,
\label{Korringa}
\end{equation} 
where $\gamma_{e}$ is the gyromagnetic ratio of electrons. 
Figure~2 shows the temperature dependence of $1/(T_1T)$ 
in Hg$_2$Ru$_2$O$_7$ at the $^{199}$Hg sites measured at the peak frequency of the spectra. 
Unlike in simple metals, $1/(T_1T)$ increases slightly with 
decreasing temperature. Such an increase in $1/(T_1T)$ has often been 
ascribed to the electron-electron interaction causing the growth 
of low-energy spin fluctuations\cite{Moriya}. For instance, $1/(T_1T)$ at the Cu sites in high-$T_c$ 
cuprate superconductors shows Curie-Weiss temperature dependence due to the 
development of antiferromagnetic spin fluctuations\cite{Millis}. When the low-frequency 
spin fluctuations are associated with spatial correlation at a particular wave 
vector ${\bf q}$, $1/(T_1T)$ and $K_{\rm spin}$ are affected in different ways. 
If ferromagnetic correlation develops near $q$ = 0, $K_{\rm spin}$ is enhanced 
much more strongly than $1/(T_1T)$, resulting in $1/(T_{1}TK_{\rm spin}^{2})$ 
becoming significantly smaller than $S$ in eq.~(\ref{Korringa}). On the other hand, 
antiferromagnetic spin fluctuations at a non-zero $q$ enhance $1/(T_1T)$, 
but do not substantially change $K_{\rm spin}$, making $1/(T_{1}TK_{\rm spin}^{2})$
much larger than $S$. 

\begin{figure}[tb]
\begin{center}
\includegraphics[width=0.9\linewidth]{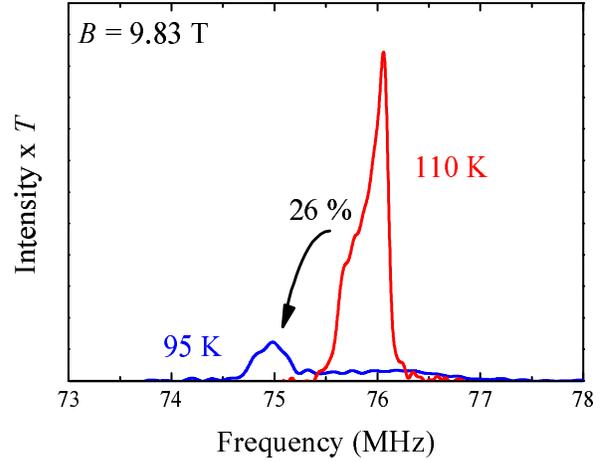}
\end{center}
\caption{(Color online) $^{199}$Hg-NMR spectra at 95 and 110 K at 9.83 T. 
The vertical axis represents the spectral intensity multiplied by temperature.}
\label{fig3}
\end{figure} 
In Fig.~2, $1/(T_{1}TK_{\rm iso}^{2})$ is plotted against temperature and 
compared with $S$ for $^{199}$Hg nuclei (the red line). 
Here, we consider only the isotropic component of $K_{\rm spin}$ due 
to $s$ electrons and neglect the much smaller axial part. 
The result that $1/(T_{1}TK_{\rm iso}^{2})$ is strictly independent 
of temperature with a value very close to $S$ rules out strong spatial spin correlation 
at any specific wave vector. On the other hand, the density of states 
at the Fermi level calculated by the local-density approximation, 
$\rho = 3.7$~states/eV-Ru\cite{Craco,Harima}, gives the bare spin susceptibility 
$\rho\mu_{B}^{2} =1.2 \times 10^{-4}$~emu/mol-Ru. Since the measured 
$\chi \sim 5.5 \times 10^{-4}$ emu/mol-Ru is dominantly due to a spin 
contribution, we conclude that $\chi_{\rm spin}$ is enhanced by a factor 
of five. This modest enhancement of $\chi_{\rm spin}$ and 
the absence of spatial spin correlation leads us to conclude that the spin 
fluctuations are local, i.e., largely uniform in ${\bf q}$-space. Such behavior 
is similar, for example, to that of the local Fermi liquid in Kondo systems\cite{Shiba}. 
We may also say that the temperature dependences of $\chi_{\rm spin}$ and 
$1(/T_1T)$ should be ascribed to the variation in the effective density of states. 
The absence of spatial correlation in  Hg$_2$Ru$_2$O$_7$ may be a manifestation 
of geometrical frustration of the pyrochlore lattice in itinerant electron systems. 

Let us now discuss the $^{199}$Hg-NMR results in the insulating phase. 
Figure 3 shows the $^{199}$Hg-NMR spectra above and below $T_{\rm MI}$. 
Significant changes in the spectral shape and intensity are observed\cite{note1}. 
The integrated intensity of the spectrum is reduced to 
26\% upon cooling across $T_{\rm MI}$, i.e., only one-quarter of the Hg sites contribute 
to the NMR signal in the frequency range between 74 and 78 MHz in the insulating phase. 
The resonance frequency of the rest of the Hg sites must be distributed over a much wider 
frequency range owing to the large hyperfine fields, suggesting a magnetically ordered state with 
spontaneous magnetic moments. 

\begin{figure}[tb]
\begin{center}
\includegraphics[width=0.9\linewidth]{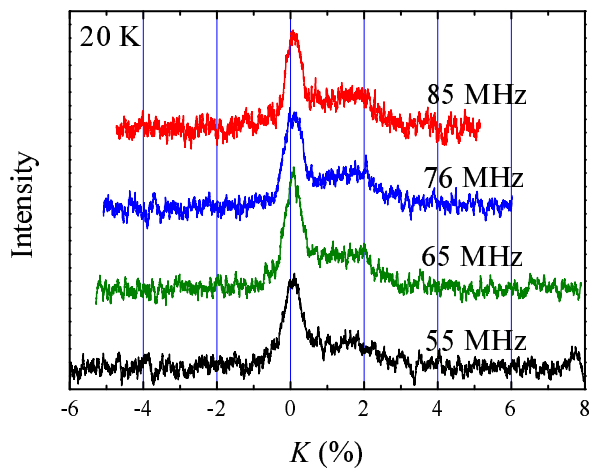}
\end{center}
\caption{(Color online) Field-swept $^{199}$Hg-NMR spectra at 
20 K obtained at different frequencies.}
\label{fig4}
\end{figure} 
Figure~4 shows the $^{199}$Hg-NMR spectra at 20 K at different frequencies. 
Here, the spectra were obtained by sweeping the magnetic field $B$ at a fixed 
frequency and plotted against the fractional shift in the resonance field, 
$K= ( \nu_{r} -~^{199}\gamma B) /^{199}\gamma B$. 
Plotted in this way, all the spectra in Fig.~4 have identical line shapes, indicating 
that the hyperfine field $\nu_{r} -~^{199}\gamma B$ is proportional to $B$. 
Therefore, for one-quarter of the Hg sites being observed, the hyperfine field from 
the spontaneous magnetic moments must be canceled out and the NMR line shape is 
determined solely by a small shift due to field-induced magnetization. 
The spectra in Fig.~4 have two components, a sharp peak at $K \sim 0$ 
and a broad structure with an edge at $K \sim 2$~\%, with nearly equal 
integrated intensities, suggesting that two types of Hg sites with equal populations 
contribute to the spectra in Fig.~4. 

If the loss of the NMR signal from three-quarters of the Hg sites is 
due to the spontaneous hyperfine field from ordered magnetic moments, we should be able to 
observe NMR signals from these Hg sites at zero magnetic field. We have searched 
zero-field NMR signals in the frequency range between 30 and 200~MHz at 
low temperatures below 4~K.  However, no signal was observed 
except for the resonance from $^{99,101}$Ru, which will be described in detail in
$\S$\ref{RuNMR}. Since the ordered moment is estimated to be 1~$\mu _B$/Ru, as
will be explained in  $\S$\ref{RuNMR}, we expect the zero-field Hg resonance to be 
in the range of 80 - 130 MHz if the moments were ferromagnetically aligned, on the 
basis of the hyperfine coupling constants determined above. However, a much smaller 
resonance frequency is expected for antiferromagnetic spin configurations due to the cancellation 
of a hyperfine field from different neighbors. Therefore, the actual resonance 
frequency in Hg$_2$Ru$_2$O$_7$ is likely to be lower than 30~MHz, 
which would be difficult to observe because of reduced NMR sensitivity  
at lower frequencies. 

\subsection{$^{99,101}$Ru-NMR} 
\label{RuNMR}

In order to obtain direct evidence for a magnetic order below $T_{\rm MI}$, 
a search was conducted for the resonance at zero magnetic field caused by a 
spontaneous hyperfine field.  We found NMR signals in the frequency 
range of 50 - 90 MHz.  The NMR spectrum at zero field obtained for 
sample $A$ with natural isotopic abundance is shown by the 
open circles in Fig.~5.  This spectrum consisting of several sharp peaks 
provides unambiguous microscopic evidence for a commensurate  
antiferromagnetic order in the insulating phase below $T_{\rm MI}$. 
\begin{figure}[tb]
\begin{center}
\includegraphics[width=0.9\linewidth]{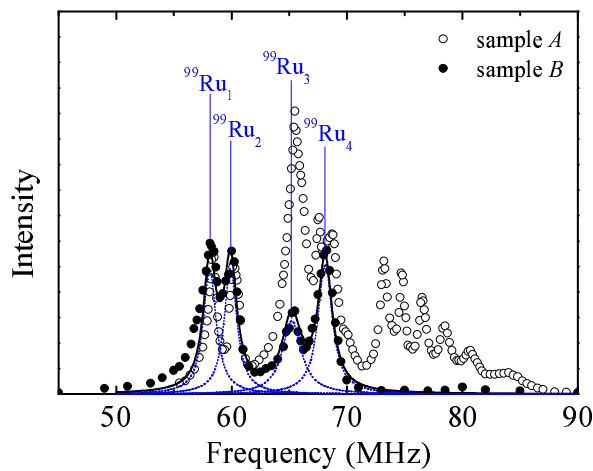}
\end{center}
\caption{(Color online) NMR spectrum at zero field obtained for 
sample $A$ with natural isotopic abundance at $T$=1.6~K (open circles) and 
sample $B$ enriched with $^{99}$Ru isotope at $T$=4.2~K (solid circles). 
The solid line represents fitting to the spectrum from sample $B$ by the sum of four 
Lorentzian functions with full widths at half maximum of 1.4, 1.4, 2.0, and 1.5~MHz for the 
$^{99}$Ru$_1$, $^{99}$Ru$_2$, $^{99}$Ru$_3$, and $^{99}$Ru$_4$ sites, 
respectively (dotted lines).}
\label{fig5}
\end{figure} 
There are three species of NMR active nuclei, $^{199}$Hg and $^{99,101}$Ru, 
in Hg$_2$Ru$_2$O$_7$ (Table I). In order to assign the resonance lines 
to specific nuclear species, we have also performed zero-field NMR measurements 
on sample $B$ enriched with the $^{99}$Ru isotope. The spectrum from 
sample $B$ is shown by the solid circles in Fig.~5. The isotopic enrichment results in a 
marked change of the spectrum. In particular, all the resonance lines above 70~MHz 
observed in sample $A$ are absent in sample $B$. Therefore, they should be assigned to 
$^{101}$Ru. 

\begin{figure*}[t]
\begin{center}
\includegraphics[width=0.9\linewidth]{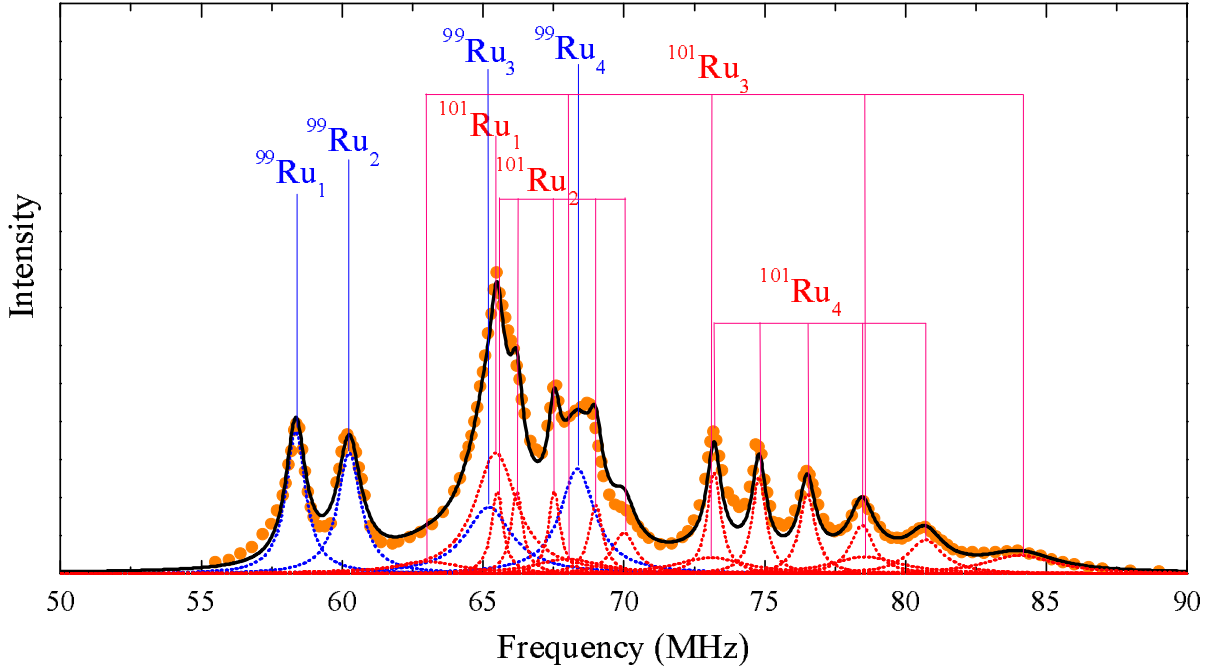}
\end{center}
\caption{(Color online) Zero-field NMR spectrum (solid circles) obtained for sample $A$ at 1.6 K is 
reproduced by the sum of Lorentzian functions (solid lines), each of which (dotted lines) represents 
the peak from $^{99}$Ru or $^{101}$Ru nuclei on four inequivalent sites with equal populations.}
\label{fig6}
\end{figure*} 
Since both $^{99}$Ru and $^{101}$Ru nuclei have spin $I$ = 5/2, the electric quadrupole 
interaction between the nuclear quadrupole moments and the electric field gradient (EFG) 
should split the zero-field resonance into five equally spaced lines up to the first-order 
perturbation in the quadrupole interaction. Since the Ru sites have the 
trigonal symmetry in Hg$_2$Ru$_2$O$_7$, the resonance frequency 
$\nu _M$ for the transition $I_{z} = M \leftrightarrow  M - 1$ is given as\cite{note2} 
\begin{equation}
\nu _M  =~^{\alpha }\gamma B_{\rm hf}^{\rm Ru} + \frac{^{\alpha}\nu _q}{2}\left( 3\cos^2\phi  - 1\right) \left( M - \frac{1}{2} \right) , 
\label{res}
\end{equation}
where $M$ takes five half-integer values between $-$3/2 and 5/2, $^{\alpha}\gamma$ is 
the nuclear gyromagnetic ratio for $^{99}$Ru ($\alpha $ = 99) or $^{101}$Ru ($\alpha$ = 101), 
$B_{\rm hf}^{\rm Ru}$ is the magnitude of the hyperfine field from the spontaneous magnetic 
moments, and $\phi $ is the angle between ${\bf B}_{\rm hf}^{\rm Ru}$ and the 
trigonal axis of the Ru site being observed. The maximum quadrupole splitting 
$^{\alpha}\nu _{q} = 3e^{\alpha }QV_{zz}/ [2I(I - 1)h]$ is determined by the 
electric field gradient $V_{zz}$ along the trigonal axis and the nuclear 
quadrupole moment $^{\alpha }Q$ of the $^{\alpha}$Ru nuclei. 

We first examine the spectrum of sample $B$.  The spectrum 
consists of four peaks, which should be assigned to $^{99}$Ru. The 
non-uniform spacing between the peaks is in clear contradiction to 
the prediction of five equally spaced lines for a given 
$B_{\rm hf}^{\rm Ru}$. We therefore conclude that the quadrupole splitting 
for $^{99}$Ru is too small to be resolved as distinct peaks. The 
peaks in the spectrum should then be assigned to four types of Ru sites 
with different values of $B_{\rm hf}^{\rm Ru}$. The spectrum from sample $B$ is 
fit to the sum of four Lorentzian functions, as shown in Fig.~5. The four sites 
are labeled $^{99}$Ru$_1$, $^{99}$Ru$_2$, $^{99}$Ru$_3$, 
and $^{99}$Ru$_4$. The unresolved quadrupole splitting $|(\nu _q/2)(3\cos^2\phi  - 1)|$ 
should contribute to the line width. $^{99}$Ru$_3$ has the 
largest width, $^{99}$Ru$_1$ and $^{99}$Ru$_2$ have 
relatively small widths. On the other hand, the integrated intensities of these lines 
are almost the same, indicating equal populations of the four sites. The hyperfine field $B_{\rm hf}^{\rm Ru}$ 
for each site was obtained by dividing the peak frequency by $^{99}\gamma$
and is listed in Table II. 

We now turn to the spectrum of sample $A$, which should contain 
resonance lines from four $^{101}$Ru sites ($^{101}$Ru$_1$, 
$^{101}$Ru$_2$, $^{101}$Ru$_3$, 
and $^{101}$Ru$_4$) in addition to the contribution of the four $^{99}$Ru sites. 
Since $^{101}$Ru nuclei have much larger quadrupole moments than $^{99}$Ru, 
$^{101}Q/^{99}Q = 5.8$, quadrupole splitting may be well resolved in the 
spectrum of $^{101}$Ru.  Since the frequency of the center line ($M$ = 1/2 in eq.~\ref{res}) 
is not affected by the quadrupole interaction up to the first order, their positions 
for the $^{101}$Ru sites are given by $^{101}\gamma B_{\rm hf}^{\rm Ru}$, where 
the values of $B_{\rm hf}^{\rm Ru}$ are listed in Table II. We have succeeded in 
reproducing the whole spectrum of sample $A$ by adjusting the quadrupole 
splitting for the four $^{101}$Ru sites, as indicated by the solid line in Fig.~6. 

The values of the quadrupole splitting for the $^{101}$Ru sites 
were determined as follows. First, one can easily recognize five lines 
with approximately uniform spacing in the frequency range of 73 - 81 MHz. 
Since the peak frequency of the center line agrees precisely with 
$B_{\rm hf}^{\rm Ru}$ at the $^{99}$Ru$_4$ site multiplied by $^{101}\gamma$, 
these five lines are assigned to the $^{101}$Ru$_4$ sites. We note, however, that 
the quadrupole splittings between adjacent lines are not completely uniform 
but take slightly different values, namely, 1.6, 1.7, 2.0, and 2.2 MHz 
from low to high frequencies. This should be due to the higher-order effects of 
the quadrupole interaction.  Next, the difference of the spectra of the two samples 
between 63 and 72~MHz is attributed to the  $^{101}$Ru$_1$ and 
$^{101}$Ru$_2$ sites, on the basis of the values of $B_{\rm hf}^{\rm Ru}$ of these sites. 
The spectral shape in this frequency range is well reproduced by assigning 
a very small unresolved quadrupole splitting 
less than 0.5~MHz to the $^{101}$Ru$_1$ sites and a barely resolved small 
splitting of 1~MHz to the $^{101}$Ru$_2$ sites, as indicated in Fig.~6. 
The small values of the quadrupole splitting for these sites are 
consistent with the small line widths of the $^{99}$Ru$_1$ and 
$^{99}$Ru$_2$ sites. Finally, a large quadrupole splitting is expected 
for the $^{101}$Ru$_3$ sites because the $^{99}$Ru$_3$ sites show 
the largest width.  There is indeed a yet unidentified broad peak 
at 84~MHz, which must be assigned to the highest frequency quadrupole 
satellite ($M$ = 5/2) for the $^{101}$Ru$_3$ sites. The quadrupole splitting 
for $^{101}$Ru$_3$ is then determined to be 5~MHz, generating other 
satellite lines.  The values of the quadrupole splitting 
for all $^{101}$Ru sites are listed in Table II. 

\begin{table}[tb]
\caption{Hyperfine field $B_{\rm hf}^{\rm Ru}$ values and the quadrupole splittings at the $^{101}$Ru sites.}
\label{HypFieldRu}
\begin{tabular}{ccc}
\hline
  & $B_{\rm hf}^{\rm Ru}$ (T) & $|(\nu _q/2)(3\cos^2\phi  - 1)|$ (MHz) \\
\cline{2-3}
Ru$_1$ & 29.8 & $\leq $ 0.5 \\
Ru$_2$ & 30.7 & 1 \\
Ru$_3$ & 33.3 & 5 \\
Ru$_4$ & 34.9 & 2 \\
\hline
\end{tabular}
\end{table} 

The above analysis of the zero-field NMR spectra indicates that all four Ru sites 
have similar magnitudes of the hyperfine field in the range of 30 - 35~T. Since the 
hyperfine field at the Ru nuclei ${\bf B}_{\rm hf}^{\rm Ru}$ mainly comes 
from the spin density of the 4$d$ states on the same sites, the magnitude of the 
ordered moments, ${\bf m}$, can be determined from the relation 
\begin{equation}
{\bf B}_{\rm hf}^{\rm Ru} = {\bf A}^{\rm Ru} \cdot {\bf m} 
\label{RuHypField}
\end{equation}
once the hyperfine coupling tensor ${\bf A}^{\rm Ru}$ is known. 
Unfortunately, we were unable to determine ${\bf A}^{\rm Ru}$ 
in Hg$_2$Ru$_2$O$_7$, because the Ru NMR signal could not be observed
in the paramagnetic phase probably owing to too short a spin-echo decay time.  
However, the hyperfine field from the 4$d$ spin is mostly due to the 
core-polarization effect, which yields an isotropic coupling constant of 
$-30 \pm  6$~T/$\mu _B$. This value is basically a single-ion property
common to all compounds\cite{Watson, Mukuda}. Using this value, the 
ordered moments in the antiferromagnetic phase are estimated 
to be in the range $1.1 \pm 0.3 \mu _B$. 

Very recently, a magnetic 
order below $T_{\rm MI}$ has also been detected by muon spin 
rotation ($\mu $SR) experiments\cite{Miyazaki}. The muon spectrum shows 
multi-components whose frequencies differ by a factor of three.  
Since the Ru-NMR results show a largely uniform magnitude of ordered moments,  
the $\mu $SR results indicate either multiple muon sites or multiple internal fields at 
one crystallographic site due to different degrees of cancellation from   
neighboring antiferromagnetic moments. The latter is the case for Hg nuclei, as we 
discussed in $\S$\ref{HgNMR}.

If we assume a Mott insulating phase below $T_{\rm MI}$ and take 
the localized ionic picture with strong correlation, three 4$d$ electrons of 
Ru$^{5+}$ occupying half of the $t_{2g}$ states should form a spin 3/2 with no 
orbital degree of freedom by the Hund's coupling. We then expect an ordered
moment of  3~$\mu _B$ per Ru, which is three times larger than the observed value. 
Neither the deviation of the crystal field from a cubic symmetry in the pyrochlore oxides 
nor the spin-orbit coupling in 4$d$-transition-metal elements is believed to be 
strong enough to break the Hund's coupling.  Therefore, it appears difficult 
to explain the small ordered moments by localized spins in the Mott's picture 
in the strong correlation limit.  Localized spin models also seem incompatible 
with the high transition temperature. An antiferromagnetic order of localized 
moments on a highly frustrated pyrochlore lattice has to be 
driven by weak secondary interactions or by partial removal of frustration due to  
structural distortion.  However, neither of these is likely to be sufficient to 
stabilize the antiferromagnetic order in Hg$_2$Ru$_2$O$_7$ up to such 
a high temperature as $T_{\rm MI}$ = 107~K. It thus appears more 
appropriate to understand the AF insulating state using itinerant electron models. 
One may argue that a spin-density-wave transition in a three-dimensional 
pyrochlore lattice is not likely to open a full gap across the entire Fermi 
surfrace.  This is true in the limit of weak correlation. We speculate, however, that 
moderate correlation effects may open a full gap in the AF state.  

The stability of the insulating state in Hg$_2$Ru$_2$O$_7$ has been examined  
theoretically by combining the local-density approximation and the dynamical mean-field 
theory by Craco \textit{et al.}\cite{Craco} They found that a significant reduction in the 
4$d$ electron number as a result of the charge transfer between Ru-O(1) and 
Hg-O(2) sublattices is necessary to stabilize a paramagnetic insulating state.  
It should be interesting to examine if the insulating state is stabilized 
without such a charge transfer in the presence of an antiferromagnetic order. 
Another point reported by Craco \textit{et al.} is the nearly localized character of 
the $a_{1g}$ state in contrast to the more itinerant $e^{\prime}_{g}$ 
carrier split by the trigonal crystal field\cite{Craco}. We note that the 
antiferromagnetic order of the $a_{1g}$ electrons alone should result in an 
ordered moment of 1~$\mu_{B}$.     

\subsection{Possible magnetic structure} 
\label{structure}

Let us now discuss the magnetic structure below $T_{\rm MI}$. 
Since spin systems on the pyrochlore lattice with the nearest-neighbor 
Heisenberg interaction alone remain disordered, magnetic orders induced 
by various secondary interactions have been studied extensively. 
Several types of magnetic structures associated with the 
wave vector ${\bf q}=0$ were found theoretically in the presence of 
$XY$-anisotropy\cite{Bramwell}, dipolar interaction\cite{Palmer}, and 
DM interaction\cite{Elhajal}. ${\bf q}=0$ magnetic structures
were actually observed by neutron scattering experiments, for example, in 
Gd$_2$Sn$_2$O$_7$\cite{Wills061} and Er$_2$Ti$_2$O$_7$\cite{Poole071}.
The structure in Gd$_2$Sn$_2$O$_7$ agrees with the theoretical prediction
for a nearest-neighbor Heisenberg antiferromagnet with dipolar interaction\cite{Palmer}. 
On the other hand, magnetic Bragg peaks with ${\bf q}$ = (1/2,  1/2,  1/2) 
were observed in Gd$_2$Ti$_2$O$_7$\cite{Champion,Stewart}.   

The notion of magnetic order stabilized by a weak secondary interaction 
is probably not relevant to Hg$_2$Ru$_2$O$_7$, since the ordered
state is stable up to such a high temperature above 100~K.  However, 
knowledge of magnetic structures should still be crucial to understand the 
mechanism of the MI transition and magnetic order. In the following, 
we discuss possible spin structures compatible with the NMR data.  
There are two major observations that put strong constraints 
on the spin structure: (1) the hyperfine field at one-quarter of 
the Hg sites is cancelled out, (2) there are four inequivalent 
Ru sites with different values of $B_{\rm hf}^{\rm Ru}$ and 
quadrupole splitting. The following discussion is based on the ideal cubic 
pyrochlore structure and we ignore the small structural distortion in the 
insulating phase unless explicitly noted otherwise. 

\begin{figure}[t]
\begin{center}
\includegraphics[width=0.5\linewidth]{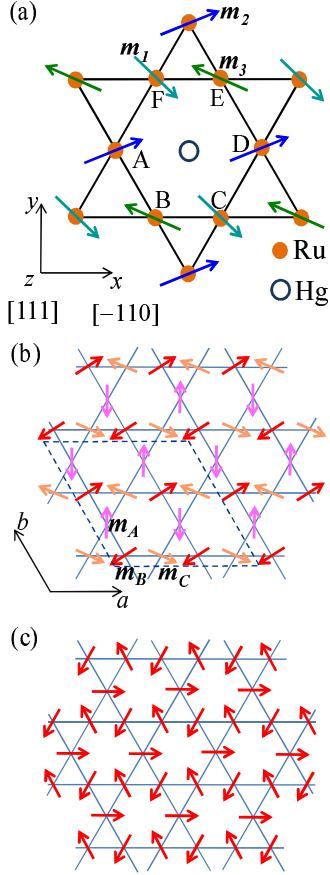}
\end{center}
\caption{(Color online) (a) The kagome layer of the Ru pyrochlore 
lattice with the central Hg site viewed down the [111] direction. An
example of the ${\bf q}=0$ spin structure is shown. 
(b) The antiferromagnetic spin structure within a kagome layer compatible with the 
NMR results. The dashed line indicates a 2$\times$2 magnetic unit cell. 
(c) The ${\bf q}=0$ spin structure within a kagome layer that 
satisfy the cancellation of the hyperfine field at the central Hg nuclei.}
\label{fig7}
\end{figure} 
We first consider condition (2). The hyperfine field at 
Ru sites is expressed as $B_{\rm hf}^{\rm Ru} = |{\bf A}^{\rm Ru} \cdot {\bf m}|$
(eq.~\ref{RuHypField}). If we take only the dominant isotropic part 
of ${\bf A}^{\rm Ru}$ due to the cope polarization effects, ${\bf B}_{\rm hf}^{\rm Ru}$ is 
parallel to ${\bf m}$ and $B_{\rm hf}^{\rm Ru}$ depends only on the magnitude of 
${\bf m}$. Here we 
consider more general case, where the anisotropic dipolar contribution is 
included in ${\bf A}^{\rm Ru}$. Because of the trigonal symmetry at the Ru sites,  
${\bf A}^{\rm Ru}$ should be axially symmetric along the $\langle 111\rangle$ 
direction. Then 
\begin{equation}
B_{\rm hf}^{\rm Ru} = m \sqrt{(A_{\parallel }^{\rm Ru})^2\cos^2\psi + (A_{\perp }^{\rm Ru})^2 \sin^2\psi }, 
\end{equation}
where $\psi$ is the angle between the trigonal axis and ${\bf m}$, and 
$A_{\parallel}^{\rm Ru}$ ($A_{\perp}^{\rm Ru}$) 
is the principal values of ${\bf A}^{\rm Ru}$ parallel 
(perpendicular) to the trigonal axis. Thus, $B_{\rm hf}^{\rm Ru}$ depends 
on both $m$ and $\psi$. The different values of $B_{\rm hf}^{\rm Ru}$ for 
the four Ru sites can be due to difference in $m$ or $\psi$ or both.  

On the other hand, the relative ratios of the quadrupole splitting 
$|\nu_{q}(3\cos^2\phi  - 1)|$ among different sites depend only on 
$\phi$ since $\nu_{q}$ is common to all Ru sites, though it is 
unknown. The very small quadrupole splitting at the Ru$_1$ 
sites (Table II) suggests that $\phi $ for Ru$_1$ is in the neighborhood 
of the \textit{magic angle}, $\phi = 54.7^{\circ }$, where 3$\cos^{2}\phi - 1 = 0$. 
On the other hand, $\phi $ for the Ru$_3$ sites with the large splitting 
should be far from 54.7$^{\circ}$.  Note that $\phi$, which is the angle 
between the trigonal axis and ${\bf B}_{\rm hf}^{\rm Ru}$, is nearly equal to 
$\psi$ since  ${\bf A}^{\rm Ru}$ is approximately isotropic. 
%It is noted that signals for antiparallel spins with $\phi $ = 0 and 180$^{\circ }$ 
%cannot be separated in NMR measurements at zero field. 

Next we consider constraints from condition (1).  In 
Hg$_2$Ru$_2$O$_7$, both Ru and Hg atoms form 
distinct pyrochlore lattices. Each pyrochlore lattice consists 
of two-dimensional kagome and triangular layers alternately 
stacked along the $\langle 111\rangle$ direction. The kagome 
layers of the Ru sublattice are coplanar with the triangular 
layers of the Hg sublattice, which contains one-quarter of the 
Hg sites. Therefore, it is natural to look for a spin structure 
in which the hyperfine field at these Hg sites vanishes. 
Each of these Hg sites has six nearest-neighbor Ru sites 
on a kagome layer, as shown in Fig.~7(a). These Ru sites 
are labeled $A$ to $F$ in Fig.~7(a). Since the hyperfine 
interaction in insulators are short-ranged, we can assume 
that the hyperfine field at the Hg sites is the sum of the contributions 
of the six nearest-neighbor Ru sites, 
\begin{equation}
{\bf B}_{\rm hf}^{\rm Hg} =  \sum_{i = A}^{F}{\bf A}_{i}^{\rm Hg} \cdot {\bf m}_{i} , 
\label{HypFieldHg}
\end{equation} 
where ${\bf A}_{i}^{\rm Hg}$ is the hyperfine coupling tensor between the
Hg nucleus at the center of the hexagon and the magnetic moment 
${\bf m}_{i}$ of the $i$-th Ru site ($i = A , \ldots , F$). 
Since the $x$-axis connecting the $A$ site and the central Hg nucleus 
has a $C_2$ symmetry, it is one of the principal axes of ${\bf A}_{A}^{\rm Hg}$, 
which can be expressed as   
\begin{equation}
{\bf A}_{A}^{\rm Hg} = \begin{pmatrix} A_{xx} & 0 & 0 \\ 0 & A_{yy} & A_{yz} \\ 0 & A_{yz} & A_{zz} \end{pmatrix} , 
\label{AA}
\end{equation}
using the $xyz$-coordinate shown in Fig.~7(a).  
The hyperfine coupling tensors for the $C$ and $E$ sites are 
obtained using successive 120 degree rotations, 
\begin{gather}
{\bf A}_C^{\rm Hg} = \begin{pmatrix} 
\frac{1}{4}A_{xx} + \frac{3}{4}A_{yy} & -\frac{\sqrt{3}}{4}(A_{xx} - A_{yy}) & -\frac{\sqrt{3}}{2}A_{yz} \\ 
-\frac{\sqrt{3}}{4}(A_{xx} - A_{yy}) & \frac{3}{4}A_{xx} + \frac{1}{4}A_{yy} & -\frac{1}{2}A_{yz} \\ 
-\frac{\sqrt{3}}{2}A_{yz} & -\frac{1}{2}A_{yz} & A_{zz} 
\end{pmatrix} , \nonumber \\
{\bf A}_E^{\rm Hg} = \begin{pmatrix} 
\frac{1}{4}A_{xx} + \frac{3}{4}A_{yy} & \frac{\sqrt{3}}{4}(A_{xx} - A_{yy}) & \frac{\sqrt{3}}{2}A_{yz} \\ 
\frac{\sqrt{3}}{4}(A_{xx} - A_{yy}) & \frac{3}{4}A_{xx} + \frac{1}{4}A_{yy} & -\frac{1}{2}A_{yz} \\ 
\frac{\sqrt{3}}{2}A_{yz} & -\frac{1}{2}A_{yz} & A_{zz} 
\end{pmatrix} . 
\label{tensor}
\end{gather}
Since pairs of sites $B$ and $E$, $D$ and $A$, and $F$ and $C$ are related 
each other by inversion with respect to the central Hg sites, their coupling tensors 
are identical; ${\bf A}_{B}^{\rm Hg} = {\bf A}_{E}^{\rm Hg}$, 
${\bf A}_{D}^{\rm Hg} = {\bf A}_{A}^{\rm Hg}$, 
and ${\bf A}_{F}^{\rm Hg} = {\bf A}_{C}^{\rm Hg}$. 

One simple way to cancel out the hyperfine field at the central 
Hg site is to place antiparallel moments with equal magnitudes 
on every pair of sites related by inversion: 
\begin{equation} 
{\bf m}_A = - {\bf m}_D, \ {\bf m}_B = - {\bf m}_E, \ {\bf m}_C = - {\bf m}_F .
\label{condition}
\end{equation}
Since the hyperfine coupling tensors 
are the same within each pair, the total hyperfine field is zero. 
An example of such a spin structure is shown in Fig.~7(b). 
Here, the magnitude and directions of the three moments, 
${\bf m}_A$, ${\bf m}_B$, and ${\bf m}_C$, in a kagome layer 
can be chosen in an arbitrary manner. In addition, the moments on the 
triangular layers of the Ru pyrochlore sublattice can still be freely choosen. 
It is obvious that this model provides sufficient freedom to reproduce 
the zero-field Ru-NMR spectrum (condition 1). Furthermore, if the
moments on the triangular layer are aligned uniformly within one 
layer and oppositely directed on the neighboring triangular layers, the 
cancellation of the hyperfine field is valid even if the transferred hyperfine 
interactions up to the third neighbors are considered. 

The spin structure of a single kagome layer considered here 
is described by multiple $q$-vectors consisting of ${\bf q}_1 = (1/2, 0)$, 
${\bf q}_2 = (0, 1/2)$, and ${\bf q}_3 = (1/2, 1/2)$, where the lattice unit vector
${\bf a} = [1, 0]$ and ${\bf b} = [0, 1]$ of the kagome lattice are defined as shown in 
Fig.~7(b). For example, the structure shown in Fig.~7(b)
is expressed as
\begin{align}
{\bf m}\left({\bf r}\right) &  =  {\bf m}_A \left[\sin \left( 2\pi {\bf Q}_{3a} \cdot {\bf r} \right)
+ \sin \left( 2\pi {\bf Q}_{3b} \cdot {\bf r} \right) \right] /2  \nonumber \\ 
& + {\bf m}_B \left[\cos \left( 2\pi {\bf Q}_{1a} \cdot {\bf r} \right)
+ \cos \left( 2\pi {\bf Q}_{1b} \cdot {\bf r} \right) \right] /2 \nonumber \\ 
& + {\bf m}_C \left[\cos \left( 2\pi {\bf Q}_{2a} \cdot {\bf r} \right)
- \cos \left( 2\pi {\bf Q}_{2b} \cdot {\bf r} \right) \right] /2 , 
\label{SpinStructure}
\end{align} 
where ${\bf Q}_{1a} = (1/2, 0)$, ${\bf Q}_{1b} = (1/2, 1)$, 
${\bf Q}_{2a} = (0, 1/2)$, ${\bf Q}_{2b} = (1, 1/2)$,
${\bf Q}_{3a} = (1/2, 1/2)$, ${\bf Q}_{3b} = (-1/2, 1/2)$. 
Of course, other structures are also possible. The condition 
eq.~(\ref{condition}) requires that the ordering wave vector of the 
A-sublattice is either ${\bf q}_1$ or ${\bf q}_2$.
Likewise, the ordering of the B (C)-sublattice must have the wave vector
${\bf q}_1$ or ${\bf q}_3$ (${\bf q}_2$ or ${\bf q}_3$ ).
In all the cases, the magnetic unit cell is 2$\times$2 times larger than the 
crystaline unit cell, as shown by the dashed line in Fig.~7(b). 
However, to our knowledge, such magnetic structures have 
not been observed in materials containing kagome
or pyrochlore lattices.  

We also examined simpler structures in a single kagome layer 
with ${\bf q}=0$, which conserve the translational symmetry of the  
kagome lattice, as shown in Fig.~7(a). In this case, the requirement 
that the hyperfine field at the central Hg site should vanish under general 
conditions (not by coincidence) completely determines the spin structure, 
as shown in Fig.~7(c). Here, the moments on a kagome layer are 
directed perpendicular to the mirror plane, forming a 120 
degree structure.  Details of this analysis are presented in the appendix. 
This is the structure proposed for the kagome layers of 
Gd$_2$Ti$_2$O$_7$ in ref.~\citen{Champion}, where the 
spins on the triangular layers are reported to remain disordered. 
We should note that this structure for a single kagome layer is 
compatible with the Bragg peak at ${\bf q}$ = (1/2,  1/2,  1/2) 
of the cubic pyrochlore structure observed in 
Gd$_2$Ti$_2$O$_7$\cite{Champion}. It should also be mentioned 
that subsequent experiments\cite{Stewart}revealed that the 
antiferromagnetic order in the ground state of Gd$_2$Ti$_2$O$_7$ 
has a 4-$k$ structure, not the single $k$ structure considered here. 

The structure shown in Fig.~7(c), however, is not compatible with 
the NMR observation of four inequivalent Ru sites, unless the lowering of the 
crystal symmetry due to structural transition is explicitly considered. 
Since all the moments on the kagome layers, ${\bf m}_1$, 
${\bf m}_2$, and ${\bf m}_3$, are perpendicular to the trigonal 
axes of their respective sites, $\phi$ = $\psi$ = 90$^{\circ}$, 
three Ru sites must have the same quadrupole splitting equal to 
$|\nu _q/2|$. Even the fourth Ru sites on 
the triangular layers have different quadrupole splitting, only two sites 
are allowed in this structure.  More quantitatively, the results in Table II 
impose $|\nu_q|$, which is the possible maximum quadrupole 
splitting, to be equal or larger than 5~MHz, which is the largest 
quadrupole splitting observed at the Ru$_3$ sites. 
Therefore, $|\nu _q/2| \geq 2.5$~MHz.  However, this value is 
considerably larger than the quadrupole splitting at any other site (see Table II). 
In principle, the structural distortion observed below $T_{\rm MI}$ 
may change the EFG and break the symmetry of the hyperfine coupling tensors shown in 
eqs.~(\ref{AA}) and (\ref{tensor}), allowing four distinct sites. 
However, the small lattice distortion observed in Hg$_2$Ru$_2$O$_7$ 
appears to be insufficient to resolve the large quantitative discrepancy. 

In the above discussion, we have not considered explicitly the presence of two 
inequivalent sites within one-quarter of the Hg sites with cancelled hyperfine fields, 
as shown in Fig.~4. Such a subtle distinction could be due to 
the structural distortion in the insulating phase. Detailed structural 
analysis would be necessary to understand the origin of the two inequivalent Hg sites. 

\section{Summary} 
We have presented the $^{199}$Hg- and $^{99,101}$Ru-NMR results 
of the pyrochlore oxide Hg$_2$Ru$_2$O$_7$. In the metallic phase 
above $T_{\rm MI}$, the nuclear spin-lattice relaxation rate 1/$T_1$ 
and the magnetic shift at the $^{199}$Hg sites follow the Korringa relation, 
which indicates the absence of spatial magnetic correlations. 
Below $T_{\rm MI}$, $^{99,101}$Ru-NMR lines are observed at zero field, 
providing evidence for a commensurate antiferromagnetic order.  
This is in contrast to the singlet ground state of another pyrochlore oxide
showing a MI transition Tl$_2$Ru$_2$O$_7$. 
The ordered moments are estimated to be about $1 \mu _B$. 
The reduced moment compared with the ionic value implies that localized 
spin models with strong Hund's coupling are inadequate. An itinerant electron 
model may be more appropriate to understand the magneic properties of Hg$_2$Ru$_2$O$_7$
even below $T_{\rm MI}$. Possible magnetic 
structures compatible with the NMR results were proposed.  It would be 
a future theoretical challenge to examine the stability of specific antiferromagnetic 
structures in the insulating state.     

\section*{Acknowledgment}

We thank Hisatomo Harima for informing us of the results of his LDA calculation.  
This work was supported by MEXT KAKENHI on Priority Areas 
``Novel State of Matter Induced by Frustration'' (Nos. 22014004 and 19052008
)  and  
JSPS KAKENHI (B) (No. 21340093) and (C) (No. 50398898).  

\appendix
\section{${\bm q}=0$ antiferromagnetic structure in a kagome layer}

We consider possible spin structures with ${\bf q}=0$ under the condition 
that a hyperfine field at Hg nuclei on the kagome layer of the Ru pyrochlore 
sublattice cancels out. A ${\bf q}=0$ structure on a kagome layer is illustrated 
in Fig. 7(a). We define the components of the sublattice moments as 
\begin{gather}
{\bf m}_C = {\bf m}_F = {\bf m}_1 = (m^1_x, m^1_y, m^1_z) \nonumber \\
{\bf m}_A = {\bf m}_D = {\bf m}_2 = (m^2_x, m^2_y, m^2_z) \nonumber \\
{\bf m}_B = {\bf m}_E = {\bf m}_3 = (m^3_x, m^3_y, m^3_z) . 
\end{gather}
The hyperfine field at the central Hg site is given by 
\begin{align}
{\bf B}_{\rm hf}^{\rm Hg} = & ({\bf A}_C^{\rm Hg} +{\bf A}_F^{\rm Hg}) \cdot {\bf m}_1 \nonumber \\
& + ~({\bf A}_A^{\rm Hg} +{\bf A}_D^{\rm Hg}) \cdot {\bf m}_2 \nonumber \\
& + ~({\bf A}_B^{\rm Hg} +{\bf A}_E^{\rm Hg}) \cdot {\bf m}_3. 
\label{sum}
\end{align}
Each component of ${\bf B}_{\rm hf}^{\rm Hg} = (B_x, B_y, B_z)$ can be written as 
\begin{align}
B_x = & \left( \frac{1}{2}m^1_x - \frac{\sqrt{3}}{2}m^1_y + 2m^2_x + \frac{1}{2}m^3_x 
+ \frac{\sqrt{3}}{2}m^3_y \right) A_{xx} \nonumber \\
&+ \left( \frac{3}{2}m^1_x + \frac{\sqrt{3}}{2}m^1_y + \frac{3}{2}m^3_x 
- \frac{\sqrt{3}}{2}m^3_y \right) A_{yy} \nonumber \\
& + \left( -\sqrt{3}m^1_z + \sqrt{3}m^3_z \right) A_{yz},\nonumber \\
B_y = & \left( -\frac{\sqrt{3}}{2}m^1_x + \frac{3}{2}m^1_y + \frac{\sqrt{3}}{2}m^3_x 
+ \frac{3}{2}m^3_y \right) A_{xx} \nonumber \\
& + \left( \frac{\sqrt{3}}{2}m^1_x + \frac{1}{2}m^1_y  + 2m^2_y - \frac{\sqrt{3}}{2}m^3_x 
+ \frac{1}{2}m^3_y \right) A_{yy} \nonumber \\
& + \left( -m^1_z + 2m^2_z - m^3_z \right) A_{yz}, \nonumber \\
B_z = & \left( -\sqrt{3}m^1_x - m^1_y + 2m^2_y + \sqrt{3}m^3_x 
- m^3_y \right) A_{yz} \nonumber \\
& + \left( 2m^1_z + 2m^2_z + 2m^3_z  \right) A_{zz}.
\label{append}
\end{align}
The condition ${\bf B}_{\rm hf}^{\rm Hg}=0$ for general values of 
$A_{xx}$, $A_{yy}$, $A_{zz}$, and $A_{yz}$ requires the following: 
\begin{gather}
\frac{1}{2}m^1_x - \frac{\sqrt{3}}{2}m^1_y + 2m^2_x + \frac{1}{2}m^3_x 
+ \frac{\sqrt{3}}{2}m^3_y = 0 \\ 
\frac{3}{2}m^1_x + \frac{\sqrt{3}}{2}m^1_y + \frac{3}{2}m^3_x 
- \frac{\sqrt{3}}{2}m^3_y = 0 \\ 
-\sqrt{3}m^1_z + \sqrt{3}m^3_z = 0 \\ 
-\frac{\sqrt{3}}{2}m^1_x + \frac{3}{2}m^1_y + \frac{\sqrt{3}}{2}m^3_x 
+ \frac{3}{2}m^3_y = 0 \\ 
\frac{\sqrt{3}}{2}m^1_x + \frac{1}{2}m^1_y  + 2m^2_y - \frac{\sqrt{3}}{2}m^3_x 
+ \frac{1}{2}m^3_y = 0, \\
-m^1_z + 2m^2_z - m^3_z = 0 \\ 
-\sqrt{3}m^1_x - m^1_y + 2m^2_y + \sqrt{3}m^3_x - m^3_y = 0 \\ 
2m^1_z + 2m^2_z + 2m^3_z = 0.
\label{moment}
\end{gather}
From eqs. (A$\cdot $ 6), (A$\cdot $ 9), and (A$\cdot $ 11), we obtain 
\begin{equation}
m^1_z = m^2_z = m^3_z = 0,
\label{zcomp}
\end{equation} 
which means that the moments should lie within the kagome layer. 

Equations.~(A$\cdot $ 7) and (A$\cdot $ 4) can be rewitten as 
\begin{equation}
-\frac{1}{2}m^1_x + \frac{\sqrt{3}}{2}m^1_y = - \frac{1}{2}m^3_x - \frac{\sqrt{3}}{2}m^3_y = m^2_x .  
\end{equation} 
Likewise, from eqs.~(A$\cdot $ 5) and (A$\cdot $ 8) we obtain
\begin{equation}
-\frac{\sqrt{3}}{2}m^1_x - \frac{1}{2}m^1_y =  \frac{\sqrt{3}}{2}m^3_x - \frac{1}{2}m^3_y = m^2_y .  
\end{equation} 
These equations can be combined in a matrix form as  
\begin{gather}
\begin{pmatrix} m^2_x \\ m^2_y \end{pmatrix} = 
\begin{pmatrix} -1/2 & \sqrt{3}/2 \\ -\sqrt{3}/2 & -1/2 \end{pmatrix}
\begin{pmatrix} m^1_x \\ m^1_y \end{pmatrix} \nonumber \\
\begin{pmatrix} m^2_x \\ m^2_y \end{pmatrix} = 
\begin{pmatrix} -1/2 & -\sqrt{3}/2 \\ \sqrt{3}/2 & -1/2 \end{pmatrix}
\begin{pmatrix} m^3_x \\ m^3_y \end{pmatrix}. \nonumber \\
\label{120degree}
\end{gather}
Thus, ${\bf m}_1$,  ${\bf m}_2$, and ${\bf m}_3$ are 
related by successive 120 degree rotations within the kagome 
layer. Finally, by combining eqs. (A$\cdot $ 10) and (A$\cdot $ 14), 
we obtain 
\begin{equation}
m^2_y = 0. 
\label{m2y}
\end{equation} 
Therefore, the spin structure is uniquely determined as illustrated in Fig. 7(c).


\begin{thebibliography}{99} %% The number "99" means that this list has more than nine items.
\bibitem{Gardner}For a review, see J. S. Gardner, K. J. P. Gingras, and J. E. Greedan: Rev. Mod. Phys. \textbf{82} (2010) 53.
\bibitem{Reimers921}J. N. Reimers: Phys. Rev. B \textbf{45} (1992) 7287. 
\bibitem{Moessner981}R. Moessner and J. T. Chalker: Phys. Rev. B \textbf{58} (1998) 12049. 
\bibitem{Moessner982}R. Moessner and J. T. Chalker: Phys. Rev. Lett. \textbf{80} (1998) 2929. 
\bibitem{Canals981}B. Canals and C. Lacroix: Phys. Rev. Lett. \textbf{80} (1998) 2933.
\bibitem{Canals001}B. Canals and C. Lacroix: Phys. Rev. B \textbf{61} (2000) 1149. 
\bibitem{Tsunetsugu}H. Tsunetsugu:  Phys. Rev. B \textbf{65} (2001) 024415. 
\bibitem{Bramwell}S. T. Bramwell and M. J. P. Gingras: J. Appl. Phys. \textbf{75} (1994) 5523.
\bibitem{Palmer}S. E. Palmer and J. T. Chalker: Phys. Rev. B \textbf{62} (2000) 488. 
\bibitem{Elhajal}M. Elhajal, B. Canals, R. Sunyer, and C. Lacroix: Phys. Rev. B \textbf{71} (2005) 094420.
\bibitem{Yamashita}Y. Yamashita and K. Ueda: Phys. Rev. Lett. \textbf{85} (2000) 4960.
\bibitem{Tchernyshyov}O. Tchernyshyov, R. Moessner, and S. L. Sondhi: Phys. Rev. B \textbf{66} (2002) 064403. 
%\bibitem{6} M. Elhajal and M. J. P. Gingras: Phys. Rev. B \textbf{70} (2004) 174426.
\bibitem{Kondo} S. Kondo, D. C. Johnston, C. A. Swenson, F. Borsa, A. V. Mahajan, L. L. Miller, T. Gu, A. I. Goldman, M. B. Maple, D. A. Gajewski, E. J. Freeman, 
N. R. Dilley, R. P. Dickey, J. Merrin, K. Kojima, G. M. Luke, Y. J. Uemura, O. Chmaissem, and J. D. Jorgensen: Phys. Rev. Lett. \textbf{78} (1997) 3729.
\bibitem{Urano} C. Urano, M. Nohara, S. Kondo, F. Sakai, H. Takagi, T. Shiraki, and T. Okubo: Phys. Rev. Lett. \textbf{85} (2000) 1052.
\bibitem{Sleight} A. W. Sleight, J. L. Gillson, j. F. Weiher, and W. Bindloss: Solid State Commun. \textbf{14} (1974) 357.  
\bibitem{Mundrus} D. Mandrus, J. R. Thompson, R. Gaal, L. Forro, J. C. Bryan, B. C. Chakoumakos, L. M. Woods, B. C. Sales, R. S. Fishman, and V. Keppens: Phys. Rev. B \textbf{63} (2001) 195104.
\bibitem{Takeda} T. Takeda, R. Kanno, Y. Kawamoto, M. Takano, F. Izumi, A. W. Sleight, and A. W. Hewat: J. Mater. Chem. \textbf{9} (1999) 215. 
\bibitem{Sakai} H. Sakai, M. Kato, K. Yoshimura, and K. Kosuge: J. Phys. Soc. Jpn. \textbf{71} (2002) 422. 
\bibitem{Lee} S. Lee, J.-G. Park, D. T. Adroja, D. Khomski, S. Streltsov, K. A. McEwen, H. Sakai, K. Yoshimura, V. I. Anisimov, D. Mori, R. Kanno, and R. Ibberson: Nat. Mater. \textbf{5} (2006) 471. 
\bibitem{Yamamoto} A. Yamamoto, P. A. Sharma, Y. Okamoto, A. Nakao, H. A. Katori, S. Niitaka, D. Hashizume, and H. Takagi: J. Phys. Soc. Jpn. \textbf{76} (2007) 043703. 
\bibitem{Klein} W. Klein, R. K. Kremer, and M. Jansen: J. Mater. Chem. \textbf{17} (2007) 1356. 
\bibitem{Takeshita}N. Takeshita, C. Terakura, Y. Tokura, A. Yamamoto, and H. Takagi: J. Phys. Soc. Jpn. \textbf{76} (2007) 063707.
\bibitem{Clarke951} W. G. Clark, M. E. Hanson, F. Lefloch, and P. Shgransan: Rev. Sci. Instrum. \textbf{66} (1995) 2453.
\bibitem{Harris}R. G. Kidd and R. J. Goodfellow: in \textit{NMR and the Periodic Table,} ed. R. K. Harris and B. E. Mann (Academic Press, London 1978). 
\bibitem{Moriya} T. Moriya: \textit{Spin Fluctuations in Itinerant Electron Magnetism}; Springer Series in Solid State Science \textbf{56} (Springer, 1985).
\bibitem{Millis} A. J. Millis, H. Monien, and D. Pines: Phys. Rev. B \textbf {42} (1990) 167. 
\bibitem{Craco} L. Craco, M. S. Laad, S. Leoni, and H. Rosner: Phys. Rev. B \textbf{79} (2009) 075125. 
\bibitem{Harima} H. Harima: private communications.
\bibitem{Shiba} H. Shiba: Prog. Theor. Phys. \textbf{54} (1975) 967.
%\bibitem{17} J. S. Lee, Y. S. Lee, K, W. Kim, T. W. Noh, J. Yu, T. Takeda, 
%and R. Kanno: Phys. Rev. B \textbf{64} (2001) 165108.
\bibitem{note1} In evaluating the integrated intensity of the spectra in Fig.~3, 
the effect of spin-echo decay is corrected by extrapolating the integrated intensity 
to $\tau = 0$, where $\tau$ is the separation between the $\pi/2$ and $\pi$ pulses in
the spin-echo pulse sequence. 
\bibitem{note2} Although the cubic symmetry of the pyrochlore structure is actually broken in the 
insulating phase of Hg$_2$Ru$_2$O$_7$, the EFG at the Ru sites should 
still be approximately axially symmetric since the structural distortion 
is quite small\cite{Yamamoto}.
\bibitem{Watson} R. E. Watson and A. J. Freeman: in \textit{Hyperfine Interaction,} 
ed. A. J. Freeman and R. B. Frankel (Academic Press, New York, 1967) Chap. 2, p. 53. 
\bibitem{Mukuda} H. Mukuda, K. Ishida, Y. Kitaoka, K. Asayama, 
R. Kanno, and M. Takano: Phys. Rev. B \textbf{60} (1999) 12279. 
\bibitem{Miyazaki} M. Miyazaki, R. Kadono, K. H. Satoh, M. Hiraishi, S. Takeshita, A. Koda, 
A. Yamamoto, and H. Takagi: Phys. Rev. B \textbf{82} (2010) 094413. 
\bibitem{Wills061}A. S. Wills, M. E. Zhitomirsky, B. Canals, J. P. Sanchez, P. Bonville, P. Dalmas de R\'{e}otier, and A. Yaouanc: J. Phys.: Condens Matter \textbf{18} (2006) L37.
\bibitem{Poole071} A. Poole, A. S. Wills, and E. Leli\'{e}vre-Berna: J. Phys.: Condens Matter \textbf{19} (2007) 452201.
\bibitem{Champion} J. D. M. Champion, A. S. Wills, T. Fennell, S. T. Bramwell, J. S. Gardner, and M. A. Green: Phys. Rev. B \textbf{64} (2001) 140407(R).
\bibitem{Stewart} J. R. Stewart, G. Ehlers, A. S. Wills, S. T. Bramwell, and J. S. Gardner: J. Phys.: Condens Matter \textbf{16} (2004) L321.
%\bibitem{20} S.-H. Lee, C. Broholm, T. H. Kim, W. Ratcliff II, 
%and S.-W. Cheong: Phys. Rev. Lett. \textbf{84} (2000) 3718. 

\end{thebibliography}
\end{document}